\documentclass{article}
\usepackage{amsfonts}
\usepackage{amssymb}
\usepackage{amsmath}
\usepackage{amsthm}
\usepackage[english]{babel}
\begin{document}

\title{ADM analysis and massive gravity}

\author{{\bf Alexey Golovnev}\\
{\small {\it Saint-Petersburg State University, high energy physics department,}}\\
{\small \it Ulyanovskaya ul., d. 1; 198504 Saint-Petersburg, Petrodvoretz; Russia}\\
{\small agolovnev@yandex.ru, \quad golovnev@hep.phys.spbu.ru}}
\date{}

\maketitle

\begin{abstract}

This is a contribution to the Proceedings of the 7th Mathematical Physics Meeting: Summer School and Conference on
Modern Mathematical Physics, held in Belgrade 09 -- 19 September 2012.

We give an easily accessible introduction to the ADM decomposition of the curvature components. After that
we review the basic problems associated with attempts of constructing a viable massive gravity theory.
And finally, we present the metric formulations of ghost-free massive gravity models, and comment on existence problem of
the matrix square root.

\end{abstract}

\section{Introduction}

General relativity is a very successful theory. We are not aware of any experimental
contradictions to its predictions whenever we are able to test it directly. However, at the
largest scales, it fails unless we are willing to introduce huge amounts of Dark Matter and
Dark Energy. The Dark Matter must be of some yet unknown mysterious nature. And the
Dark Energy might be a mere cosmological constant, but of an overwhelmingly unnatural
value. At the end of the day, it pushes us towards reconsidering the basic foundations
of the theory of gravitational interactions.

Modifying the General Relativity, even if just by giving the graviton a mass, 
is not an easy game to play \cite{Kurt}. And it is only very recently that we
have got a decent hope \cite{dRG,dRGT} of obtaining a viable model of massive gravity.
In this contribution we review some non-perturbative aspects of this theory from the
viewpoint of the Hamiltonian ADM analysis.

\section{Review of the ADM formalism}

In the ADM approach, we use the following $(3+1)$-decomposition of the metric \cite{ADM}
\begin{equation}
\label{ADM metric}
ds^2=-\left(N^2-N_i N^i \right)dt^2+2N_i dx^i dt+\gamma_{ij}dx^i dx^j
\end{equation}
where $N$ and $N_i$ are commonly called lapse and shifts respectively, and the spatial indices are
raised and lowered by the spatial metric $\gamma$. Equivalently, $g_{00}=-\left(N^2-N_i N^i \right)$, $g_{0i}=N_i$ and $g_{ij}=\gamma_{ij}$.
And it is fairly straightforward to check that the inverse metric components are
$g^{00}=-\frac{1}{N^2}$, $g^{0i}=\frac{N^i}{N^2}$ and $g^{ij}=\gamma^{ij}-\frac{N^i N^j}{N^2}$.
Comparing this result with the Cramer's rule for $g^{00}$ we get the useful relation for
determinants, $\sqrt{-g}=N\sqrt{\gamma}$.

Let us also introduce a unit normal vector 
$n^{\mu}\equiv\left(\frac{1}{N}, -\frac{N^i}{N}\right)$ which corresponds to
the one-form $n_{\mu}\equiv\left(-N, {\overrightarrow 0}\right)$ nullifying the tangent vectors. Generically, these
objects are not covariantly constant, and one can define the extrinsic curvatures
of the spatial slices $t=const$:
\begin{equation}
\label{extrcurvdef}
K_{ij}\equiv -\bigtriangledown_i n_j=\Gamma^{\mu}_{ij}n_{\mu}=-N\Gamma^0_{ij}
\end{equation}
with covariant derivatives $\bigtriangledown_{\mu}n_{\nu}\equiv\partial_{\mu}n_{\nu}-\Gamma^{\alpha}_{\mu\nu}n_{\alpha}$ 
and Christoffel symbols 
$$\Gamma^{\alpha}_{\mu\nu}=g^{\alpha\beta}\Gamma_{\beta\mu\nu}=\frac12 g^{\alpha\beta}
\left(\partial_{\mu}g_{\beta\nu}+\partial_{\nu}g_{\beta\mu}-\partial_{\beta}g_{\mu\nu}\right)$$ of the metric $g$. 

We easily get $\Gamma_{ijk}={\mathop{\Gamma}\limits^{(3)}}\vphantom{\Gamma}_{ijk}$
where $\mathop{\Gamma}\limits^{(3)}$ denotes thel Christoffel symbols for the metric $\gamma$, and
 $\Gamma_{0ij}=\frac12 \left(\partial_i N_j+
\partial_j N_i - \dot{\gamma}_{ij}\right)$ which gives 
\begin{equation}
\label{extrcurv}
K_{ij}=-N\Gamma^{0}_{ij}=\frac{1}{2N} \left({\mathop{\bigtriangledown}\limits^{(3)}}\vphantom{\Gamma}_i N_j+
{\mathop{\bigtriangledown}\limits^{(3)}}\vphantom{\Gamma}_j N_i - \dot{\gamma}_{ij}\right)
\end{equation}
for the extrinsic curvatures (\ref{extrcurvdef}) where the three-dimensional covariant derivatives are taken 
with respect to $\gamma$. Incidentally, we note that
\begin{equation}
\label{inverse}
g^{-1}=\left(\begin{array}{cc}0 & 0\\ 0 & \gamma^{-1}\end{array}\right)-n\times n^T.
\end{equation}

In order to analyse a gravitational theory, we need to compute the curvature components
\begin{equation}
\label{Riemann}
{R^{\mu}}_{\nu\alpha\beta}=\partial_{\alpha}\Gamma^{\mu}_{\beta\nu}-\partial_{\beta}\Gamma^{\mu}_{\alpha\nu}+
\Gamma^{\mu}_{\alpha\rho}\Gamma^{\rho}_{\beta\nu}-\Gamma^{\mu}_{\beta\rho}\Gamma^{\rho}_{\alpha\nu}
\end{equation}
in terms of the ADM metric decomposition (\ref{ADM metric}). For the Levi-Civita connection we have the
symmetry properties $R_{\alpha\beta\mu\nu}=-R_{\beta\alpha\mu\nu}=-R_{\alpha\beta\nu\mu}=R_{\mu\nu\alpha\beta}$, 
and therefore need to know only three types of components $R_{ijkl}$, $R_{0ijk}$ and $R_{0i0j}$. 
The calculations are rarely presented in an
explicit and detailed form. One can either use the geometric meaning of the curvature tensor and play with
parallelly transporting normal and tangent vectors, or one can follow more direct derivations in a somewhat involved algebraic disguise in
mathematical textbooks. We want to present a straightforward, almost brute force, computation. Our main point is that it can also become rather simple
if cleverly done.

\subsection{Connection components}

It is reasonable to calculate the connection components first. Anyway, they might be needed for the matter part of
the action. We will thoroughly trade $\dot\gamma$ for the extrinsic curvatures using (\ref{extrcurv}). In
the following we need:
\begin{eqnarray}
\label{connection1}
\Gamma_{ij0}=\Gamma_{i0j} &= & -NK_{ij}+{\mathop{\bigtriangledown}\limits^{(3)}}\vphantom{\Gamma}_j N_i,\\
\Gamma_{ijk} & = & {\mathop{\Gamma}\limits^{(3)}}\vphantom{\Gamma}_{ijk},\\
\Gamma^0_{00} & = & \frac1N\left(\dot{N}+N^i \partial_i N- N^i N^j K_{ij}\right),\\
\Gamma^0_{0i}=\Gamma^0_{i0}  & = & \frac1N \left(\partial_i N - N^j K_{ij}\right),\\
\Gamma^i_{0j}=\Gamma^i_{j0} & = & -\frac{N^i \partial_j N}{N} - N\left(\gamma^{ik}-\frac{N^i N^k}{N^2}\right)K_{kj}+{\mathop{\bigtriangledown}\limits^{(3)}}\vphantom{\Gamma}_j N^i,\\
\Gamma^0_{ij} & = & -\frac{1}{N}K_{ij},\\
\label{connection2}
\Gamma^i_{jk} & = & {\mathop{\Gamma}\limits^{(3)}}\vphantom{\Gamma}^i_{jk}+ \frac{N^i}{N}K_{jk}.
\end{eqnarray}

\subsection{Riemann tensor components}

Using the formulae (\ref{connection1}) -- (\ref{connection2}), we readily obtain the curvature tensor (\ref{Riemann}). First,
\begin{multline}
\label{ijkl}
R_{ijkl}=g_{i\rho}\partial_k\Gamma^{\rho}_{lj}-g_{i\rho}\partial_l\Gamma^{\rho}_{kj}+\Gamma_{ik\rho}\Gamma^{\rho}_{lj}-\Gamma_{il\rho}\Gamma^{\rho}_{kj}\\
=-N_i\partial_k\left(\frac1N K_{jl}\right)+\gamma_{im}\partial_k\left( {\mathop{\Gamma}\limits^{(3)}}\vphantom{\Gamma}^m_{jl}
+\frac{N^m}{N}K_{jl}\right)-
\frac1N K_{jl}\left(-NK_{ik}+{\mathop{\bigtriangledown}\limits^{(3)}}\vphantom{\Gamma}_k N_i\right)\\
+{\mathop{\Gamma}\limits^{(3)}}\vphantom{\Gamma}_{ikm}\left({\mathop{\Gamma}\limits^{(3)}}\vphantom{\Gamma}^m_{lj}+ \frac{N^m}{N}K_{lj}\right)-
\left(k \leftrightarrow l\right)\\
={\mathop{R}\limits^{(3)}}\vphantom{R}_{ijkl}+K_{ik}K_{jl}-K_{il}K_{jk}
\end{multline}
Then it would be easier to calculate 
\begin{equation}
\label{trick}
n_{\mu}{R^{\mu}}_{i\alpha j}=n^{\mu}R_{\mu i\alpha j}=
\frac1N R_{0 i\alpha j}-\frac{N^k}{N}R_{k i\alpha j}
\end{equation}
 instead of $R_{0 i\alpha j}$. Indeed, we get
\begin{multline}
\label{0ijk}
n_{\mu}{R^{\mu}}_{ijk}=-N\left(\partial_j\Gamma^{0}_{ki}+\Gamma^{0}_{j\rho}\Gamma^{\rho}_{ki}\right)+N\left(\partial_k\Gamma^{0}_{ji}+\Gamma^{0}_{k\rho}\Gamma^{\rho}_{ji}\right)=
\partial_j K_{ki}+{\mathop{\Gamma}\limits^{(3)}}\vphantom{\Gamma}^m_{ki}K_{jm} -\left(j\leftrightarrow k\right) \\=
{\mathop{\bigtriangledown}\limits^{(3)}}\vphantom{\Gamma}_j K_{ki} - {\mathop{\bigtriangledown}\limits^{(3)}}\vphantom{\Gamma}_k K_{ji}
\end{multline}
Finally, after just a little bit of very simple algebra, we find
$$n_{\mu}{R^{\mu}}_{i0j}=\dot{K}_{ij}+{\mathop{\bigtriangledown}\limits^{(3)}}\vphantom{\Gamma}_i {\mathop{\bigtriangledown}\limits^{(3)}}\vphantom{\Gamma}_j N+
N{K_i}^k K_{kj}-{\mathop{\bigtriangledown}\limits^{(3)}}\vphantom{\Gamma}_j\left(K_{ik}N^k\right)-K_{kj}{\mathop{\bigtriangledown}\limits^{(3)}}\vphantom{\Gamma}_i N^k.$$
Using (\ref{0ijk}) and (\ref{trick}) we trivially transform it to a more symmetric form
\begin{equation}
\label{0i0j}
n^{\mu}n^{\nu}R_{\mu i\nu j}=\frac1N\left(\dot{K}_{ij}+
{\mathop{\bigtriangledown}\limits^{(3)}}\vphantom{\Gamma}_i {\mathop{\bigtriangledown}\limits^{(3)}}\vphantom{\Gamma}_j N+
N{K_i}^k K_{kj}-{\mathcal Lie}_{\overrightarrow N}K_{ij}\right)
\end{equation}
where ${\mathcal Lie}_{\overrightarrow N}K_{ij}\equiv N^k\partial_k K_{ij}+K_{ik}\partial_j N^k+K_{jk}\partial_i N^k$, and
the partial derivatives can be substituted by the covariant ones.

\subsection{Scalar curvature and Einstein-Hilbert action}

The formulae  (\ref{ijkl}) -- (\ref{0i0j}) represent the full Riemann tensor for the Levi-Civita connection.
Now the Ricci tensor $R_{\mu\nu}\equiv{R^{\alpha}}_{\mu\alpha\nu}$ components $R_{ij}$, $n^{\mu}R_{\mu i}$ and
$n^{\mu}n^{\nu}R_{\mu\nu}$ can be easily obtained. We skip this point and find the Ricci scalar $R\equiv g^{\mu\nu}R_{\mu\nu}$ directly.
Using (\ref{inverse}) and symmetry properties of the Riemann tensor, and the obvious relation $\gamma^{ij}\dot{K}_{ij}=
\partial_0 K^i_i+K^{ij}\dot{\gamma}_{ij}$ together with $(\ref{extrcurv})$ we get
\begin{multline}
\label{scalar}
R=g^{\mu\nu}g^{\alpha\beta}R_{\mu\alpha\nu\beta}=\gamma^{ik}\gamma^{jl}R_{ijkl}-2n^{\mu}n^{\nu}\gamma^{ij}R_{\mu i\nu j}\\
={\mathop{R}\limits^{(3)}}+K^i_i K^j_j-3K^{ij}K_{ij}-\frac2N \gamma^{ij}\dot{K}_{ij}
+\frac4N K^{ij}{\mathop{\bigtriangledown}\limits^{(3)}}\vphantom{\Gamma}_j N_i+2 \frac{N^j}{N}{\mathop{\bigtriangledown}\limits^{(3)}}\vphantom{\Gamma}_j K^i_i
-\frac2N {\mathop{\bigtriangleup}\limits^{(3)}} N\\
={\mathop{R}\limits^{(3)}}+K^{ij}K_{ij}+K^i_i K^j_j-\frac2N \dot{K}^i_i+2\frac{N^j}{N}{\mathop{\bigtriangledown}\limits^{(3)}}\vphantom{\Gamma}_j K^i_i
-\frac2N {\mathop{\bigtriangleup}\limits^{(3)}} N
\end{multline}

As a final step, let us put the Einstein-Hilbert density $\sqrt{-g}R=\sqrt{\gamma}NR$ 
into a convenient form. We use the relation 
$\partial_0 \sqrt{\gamma}=\frac{\sqrt{\gamma}}{2}\gamma^{ij}\dot{\gamma}_{ij}$ and exclude 
$\dot{\gamma}$ by virtue of (\ref{extrcurv}):
\begin{equation*}
\sqrt{-g}R=\sqrt{\gamma}N\left({\mathop{R}\limits^{(3)}}+K^{ij}K_{ij}-K^i_i K^j_j\right)
-2\sqrt{-g}\bigtriangledown_{\mu} \left(K^i_i n^{\mu}\right)
-2\sqrt{\gamma}{\mathop{\bigtriangleup}\limits^{(3)}} N
\end{equation*}
where $-2\sqrt{-g}\bigtriangledown_{\mu} \left(K^i_i n^{\mu}\right)\equiv
-2\partial_0\left(\sqrt{\gamma}K^i_i\right)+2\sqrt{\gamma}{\mathop{\bigtriangledown}\limits^{(3)}}\vphantom{\Gamma}_j \left(K^i_i N^j\right)$.

Neglecting total time derivative and covariant divergence terms, we obtain the Einsten-Hilbert action in the ADM formalism:
\begin{equation}
\label{EH}
S=\int dt d^3 x \sqrt{\gamma}N\left({\mathop{R}\limits^{(3)}}+K^{ij}K_{ij}-K^i_i K^j_j\right).
\end{equation} 
The action (\ref{EH}) would be enough for us as the kinetic part of the massive gravity models. However, as we have computed all
the curvature components, these results can be used for more complicated modified gravity models, too. Note also that the curvature tensor
in theories with non-metric connection can be viewed as an (exact at second order) variation of the Riemannian one with
respect to the connection being varied from its Levi-Civita value by non-metricity and contortion tensors.

\section{Massive gravity and its problems}

Now we come to massive deformations of general relativity. Note that only the six $\gamma_{ij}$ variables
are dynamical in the action (\ref{EH}). The lapse and shifts play the role of Lagrange multipliers which
enforce four first-class constraints corresponding to the diffeomorphism invariance of the theory. The
total number of physical degrees of freedom is two. A generic mass term would break the
gauge invariance and produce a model with six independent degrees of freedom. However, a massive spin-two field
should posess only five of them. Therefore, we have an extra scalar. 

That this scalar might be problematic, can be seen already at the level of quadratic perturbations around
Minkowski space. For linearised theory, the  Einstein-Hilbert action acquires the form of
$$-\frac14(\partial_{\alpha}h_{\mu\nu})(\partial^{\alpha}h^{\mu\nu})+
\frac12(\partial^{\alpha}h_{\mu\nu})(\partial^{\nu}h^{\mu}_{\alpha})-
\frac12(\partial_{\alpha}h^{\alpha\mu})(\partial_{\mu}h^{\beta}_{\beta})+
\frac14(\partial_{\mu}h^{\alpha}_{\alpha})(\partial^{\mu}h^{\beta}_{\beta})$$
which can be studied in the standard parametrisation of $h_{00}=2\phi$, $h_{0i}
=\partial_i b+s_i$ with $\partial_i s_i=0$, and $h_{ij}=2\psi\delta_{ij}+2\partial^2_{ij}\sigma+
\partial_i v_j+\partial_j v_i+h^{(TT)}_{ij}$ with $\partial_i v_i=0$, $\partial_i h^{(TT)}_{ij}=0$
and $h^{(TT)}_{ii}=0$:
$$-\frac14 (\partial_{\alpha}h^{(TT)}_{ij})(\partial^{\alpha}h^{(TT)}_{ij})
+\frac12\left(\partial_j\left({\dot v}_i-s_i\right)\right)^2-6{\dot\psi}^2+2(\partial_i \psi)^2+
4\psi\bigtriangleup\left(\phi-{\dot b}+{\ddot\sigma}\right)$$
where we have neglected all total derivative terms. One can see that there are two gauge invariant
variables $h^{(TT)}_{ij}$ in helicity-two (transverse traceless)
sector, two gauge invariant variables $\dot{v}_i-s_i$ in helicity-one sector, 
and two gauge invariant variables, $\psi$ and $\phi-{\dot b}+{\ddot\sigma}$, in helicity-zero 
sector. Of course, these are just the standard gauge invariant variables of the
cosmological perturbation theory in the limit of vanishing Hubble constant. Only the helicity-two
modes are physical, the others are constrained. And the latter is extremely good because otherwise
we would have got severe problems with the wrong-sign kinetric term of $\psi$.
Unfortunately, we do indeed get them once the gauge invariance is broken by, say,  a mass term.

In general, we can think of two types of a mass term, $h_{\mu\nu}h^{\mu\nu}$ and
$h^{\mu}_{\mu}h^{\nu}_{\nu}$. Both of them contain $h_{00}$ and ${h_{0i}}$ fields non-linearly (quadratically) which implies
that the corresponding Einstein equations cease to provide constraints for the spatial sector. However, there is
one particular combination of them, 
$$m^2\left(h_{\mu\nu}h^{\mu\nu}-h^{\mu}_{\mu}h^{\nu}_{\nu}\right)$$ 
discovered
by Fierz and Pauli \cite{FP},
from which $h_{00}^2$ drops. The resulting constraint kills the unwanted sixth degree of freedom and gives
a ghost-free massive gravity in the linear approximation.

Unfortunately, even in the vanishing mass limit, the linear
theory anyway contradicts observations \cite{vDV} due to the scalar graviton which couples to dust modifying the
effective gravitational constant, but not to radiation keeping the bending of light intact. It was later argued
by Vainshtein \cite{Vain} that non-linear effects will take over at small scales and restore the GR limit.
But, almost at the same time, Boulware and Deser have shown \cite{BD} that the sixth
degree of freedom comes back at the non-linear level reintroducing the ghost mode. And therefore, a
stable theory of massive gravity is probably not possible at all.

\subsection{de Rham-Gabadadze-Tolley model}

Recently, a potentially ghost-free non-linear construction of massive gravity was found \cite{dRG} in the
decoupling limit ($m\to 0$ and $M_{Pl}\to\infty$ with $m^2M_{Pl}$ fixed) by carefully
getting rid off the sixth mode at every order of perturbation theory starting from the Fierz-Pauli term. The model can be resummed
\cite{dRGT, HR1} to a deceptively simple potential
\begin{equation}
\label{potential}
V=2m^2\left( {\rm Tr}\sqrt{g^{-1}f}-3\right)
\end{equation}
around any background metric $f_{\mu\nu}$. Up to the $-3$ term which is needed to avoid
a linear in $h\equiv g-f$ contribution to the action, this is the first symmetric polynomial of eigenvalues of the
square-root matrix $\sqrt{g^{-1}f}$. 
(Interestingly, it is noted in \cite{Deser} that this
model was known to Wess and Zumino as early as in 1970.)
The other two suitable potentials are the second
$\left({\rm Tr} \sqrt{g^{-1}f}\right)^2-{\rm Tr} \left(\sqrt{g^{-1}f}\right)^2$ and the third
$\left({\rm Tr} \sqrt{g^{-1}f}\right)^3-3\left({\rm Tr} \sqrt{g^{-1}f}\right){\rm Tr} \left(\sqrt{g^{-1}f}\right)^2
+2{\rm Tr} \left(\sqrt{g^{-1}f}\right)^3$ symmetric polynomials.

\section{The Hamiltonian analysis}

In order to count the degrees of freedom non-perturbatively, one has to perform
the Hamiltonian analysis of the action (\ref{EH}) with the potential term $-\sqrt{\gamma}NV$
given by (\ref{potential}) in the case of the minimal dRGT model. We find the momenta
$\pi^{ij}\equiv\frac{\partial{\mathcal L}}{\partial{\dot\gamma}_{ij}}=\sqrt{\gamma}\left(K^k_k \gamma^{ij}-K^{ij}\right)$,
the primary constraints $\pi_N=\pi_{N_{i}}=0$, and the Hamiltonian
\begin{equation*}
H=-\int d^3 x\sqrt{\gamma}\left(N\left(\mathop{R}\limits^{({\mathit 3})}+\frac{1}{\gamma}\left(\frac12 \left(\pi^j_j\right)^2-\pi_{ik}\pi^{ik}\right)
-V\right)+2N^i\mathop{{\bigtriangledown}^k}\limits^{({\mathit 3})}\pi_{ik} \right).
\end{equation*}
For a general potential, the lapse and shifts enter the Hamiltonian non-linearly, and therefore, the secondary constraints
(obtained, as usual, by commutation of the Hamiltonian with the primary constraints) directly determine their values
instead of constraining the six spatial degrees of freedom. The major claim is that for the dRGT potentials
there is a combination of the lapse and shifts which continues to serve as a Lagrange multiplier enforcing
a constraint equation on the spatial sector.

\subsection{Hassan-Rosen proof}

Hassan and Rosen have found \cite{HR2, HR3} a redefinition of shifts
$$N^i=(\delta^i_j+ND^i_j(\gamma, n))n^j$$
such that, for the new shifts $n^i$, the square-root matrix takes the form
\begin{equation}
\label{decomposition}
\sqrt{g^{-1}f}=
\frac{1}{N\sqrt{1-n^kn^k}}
\left( \begin{array}{cc}
1 & n^i \\
-n^j & -n^i n^j 
\end{array} \right)+
\left( \begin{array}{cc}
0 & 0 \\
0 & X^{ij}(\gamma,n) 
\end{array} \right)
\end{equation}
which means that the lapse assumes the role of the Lagrange multiplier and imposes
a non-trivial constraint on the spatial sector. One can check that the same is true for the higher ghost-free potentials
because the first matrix in (\ref{decomposition}) preserves its form under exponentiation to an integer positive power.
A careful investigation shows \cite{HR5} that the subsequent consistency conditions provide the second spatial
sector constraint to form a non-degenerate pair of second class constraints on $\gamma_{ij}$, and finally, an equation
for determining the lapse. And the same is true \cite{HR4} for bimetric versions (with an independent Einstein-Hilbert term for $f_{\mu\nu}$), too.

\subsection{Auxiliary variables}

One can also approach the problem in a different manner. We can introduce a matrix $\Phi$
of auxiliary fields with a constraint that $\Phi^2=N^2g^{-1}f$ which amounts to
$$NV=2m^2\Phi^{\mu}_{\mu}+\kappa^{\mu}_{\nu}\left(\Phi^{\nu}_{\alpha}\Phi^{\alpha}_{\mu}-N^2 g^{\nu\alpha}f_{\alpha\mu}\right)$$
and, varying with respect to $\Phi$, integrate out the $\kappa$-s:
$$NV=m^2\left( \Phi^{\mu}_{\mu}+ \left(\Phi^{-1}\right)^{\mu}_{\nu}N^2 g^{\nu\alpha}f_{\alpha\mu}\right).$$
In this method it is not necessary to calculate the matrix square root explicitly, and the procedure is standard and elementary.
Unfortunately, the calculations are cumbersome. But one can easily find \cite{me} a direction in the space of must-be unphysical
variables ($N$, $N^i$, $\Phi$) along which there is no restriction at the level of secondary constraints which
proves that the number of degrees of freedom is less than six. Note that, in this construction, the equations of motion
automatically demand that a real square-root matrix $\sqrt{g^{-1}}f$ does exist.

\section{Existence of the square root}

Non-perturbative existence of the real
square-root is
one of the very interesting issues about the dRGT mode \cite{Cedric}. In the cited paper the reader can find the necessary and sufficient condition
for it: if there is a real negative eigenvalue of the $g^{-1}f$ matrix, then
it must be accompanied by even numbers of identical Jordan blocks. However, let us note that if $g^{-1}f$ has
a negative eigenvalue $-\lambda$, then the corresponding eigenvector lies in the kernel
of $\lambda g+f$ matrix. In other words, there is a degenerate linear combination of $g$ and $f$ with
positive coefficients. Even though it is anyway undesirable to couple matter to a non-trivial linear combination of metrics, one might consider
imposing a condition of non-degeneracy of all positive linear combinations. On the other hand, in full quantum gravity it might appear
to be not so much non-sensical to consider complex potentials.

We would point out that there is also a uniqueness problem for the squre root. Even the $2\times 2$ unit matrix
has an infinity of real square roots, for example:
$\left(\frac15 \left( \begin{array}{cc}
3 & -4 \\
-4 & -3
\end{array} \right)\right)^2=
\left( \begin{array}{cc}
1 & 0 \\
0 & 1 
\end{array} \right)$.
Note that this one has zero trace which shifts the vacuum energy with respect to the trivial square root. It might play an important role
in a fully quantum regime.

{\bf In conclusion}, let us mention that there are lots of other problems to take care of. Some of remaining five modes
may become ghosts in certain regimes, and there can be strong coupling issues in some
physically relevant solutions. And there are interesting problems concerning multimetric theories.
Moreover, we have to accept some form of acausality in our physical world since the superliminal propagation was
shown to be quite generic in the models of dRGT type \cite{Deser}. 
However, it is very remarkable that, in the first place, we do have a reasonable candidate for a viable massive, or bimetric, gravity
model.

{\bf Acknowledgements.} The Author is supported by Russian Foundation for Basic Research Grant No. 12-02-31214.
It is also of a great pleasure to thank the orginisers of the 7th Mathematical Physics Meeting: Summer School and Conference on
Modern Mathematical Physics for the opportunity to participate in this wonderful scientific event.

\end{document}